\documentstyle[preprint,tighten,aps]{revtex}

\begin{document}
\draft
\preprint{IFUP-TH 70/96}
\title{
Critical limit and anisotropy in the two-point correlation function
of three-dimensional $\protect\bbox{{\rm O}(N)}$ models.}
\author{Massimo Campostrini, Andrea Pelissetto, Paolo Rossi, 
and Ettore Vicari}
\address{Dipartimento di Fisica dell'Universit\`a 
and I.N.F.N., I-56126 Pisa, Italy}

\date{\today}

\maketitle

\begin{abstract}
In three-dimensional ${\rm O}(N)$ models, we investigate the
low-momentum behavior of the two-point Green's function $G(x)$ in the
critical region of the symmetric phase.  We consider physical systems
whose criticality is characterized by a a rotationally-invariant fixed
point.

In non rotationally-invariant physical systems with ${\rm
O}(N)$-invariant interactions, the vanishing of anisotropy in
approaching the rotationally-invariant fixed point is described by a
critical exponent $\rho$, which is universal and is related to the
leading irrelevant operator breaking rotational invariance. At
$N=\infty$ one finds $\rho=2$.  $1/N$ expansion and strong-coupling
calculations show that, for all values of $N\geq 0$, $\rho\simeq 2$.

Non-Gaussian corrections to the universal low-momentum behavior of
$G(x)$ are evaluated, and found to be very small.
\end{abstract}


\narrowtext

Three-dimensional ${\rm O}(N)$ models describe many important critical
phenomena in nature. We just mention that the case $N=3$ describes the
critical properties of ferromagnetic materials.  The case $N=2$ is
related to the helium superfluid transition.  The case $N=1$ (i.e.,
Ising-like systems) describes liquid-vapour transitions in classical
fluids or critical binary fluids. Finally, the limit $N\rightarrow 0$
is related to dilute polymers.

The critical behavior of the two-point correlation function $G(x)$ of
the order parameter is related to the phenomenon of critical
scattering observed in many experiments, e.g.\ neutron scattering in
ferromagnetic materials, light and X-rays in liquid-gas systems.  In
Born's approximation, the cross section $\Gamma_{fi}$ for incoming
particles (i.e.\ neutrons or photons) of momentum $p_i$ and final
outgoing momentum $p_f$ is proportional to the component $k=p_f-p_i$
of the Fourier transform of $G(x)$
\begin{equation}
\Gamma_{fi} \propto \widetilde{G}(k), \qquad k = p_f-p_i.
\label{e1}
\end{equation} 
As a consequence of the critical
behavior of the two-point function $G(x)$ at $T_c$, which is 
\begin{equation}
\widetilde{G}(k )\sim {1\over  k^{2-\eta}},
\label{e2}
\end{equation}
the cross section for $k\rightarrow 0$ (forward scattering) is
observed to diverge as $T\rightarrow T_c$.  When strictly at
criticality, the relation (\ref{e2}) holds at all momentum scales
$k \ll \Lambda$, where $\Lambda$ is a generic cut-off related to the
microscopic structure of the statystical system.  In the vicinity of
the critical point, where the relevant correlation length $\xi$ is
large but finite, the behavior (\ref{e2}) occurs for 
$\Lambda \gg k\gg 1/\xi$.

We will specifically consider systems with an ${\rm O}(N)$-invariant
Hamiltonian in the symmetric phase, i.e.\ where the ${\rm O}(N)$
symmetry is unbroken.  Furthermore, we will only consider systems with
a rotationally-symmetric fixed point.  Interesting members of this
class are systems defined on highly symmetric lattices, i.e.\ Bravais
or two-point base lattices with a tetrahedral or larger discrete
rotational symmetry.

Let us consider the second moment correlation length 
\begin{equation}
\xi_G^2 = {1\over 6}\,
   {\sum_x |x|^2 G(x)\over \sum_x  G(x)},
\label{e3}
\end{equation}
and the mass scale $M_G\equiv 1/\xi_G$.
In the critical region of the symmetric phase and at low momentum,
experiments show that $G(x)$  is well approximated by a Gaussian
behavior, i.e.
\begin{equation}
{\widetilde{G}(0)\over\widetilde{G}(k)}
\simeq 1+ {k^2\over M_G^2}.
\label{e4}
\end{equation}

Our aim is to estimate the deviations from Eq.~(\ref{e4}) 
in the critical region of the symmetric phase, 
i.e.\ for $0<T/T_c - 1\ll 1$, and in the low-momentum regime
$k^2\lesssim M_G^2$.
We focus on two quite different sources of deviations:

(i) Scaling corrections to Eq.~(\ref{e4}), depending
on the ratio $k^2/M_G^2$, and reflecting the non-Gaussian
nature of the fixed point. 

(ii) Non rotationally-invariant scaling violations, reflecting a
microscopic anisotropy in the space distribution of the spins
(assuming that no anisotropy is generated by their interaction). This
phenomenon may be relevant, for example, in the study of ferromagnetic
materials, where the atoms lie on the sites of a lattice, and
anisotropy may be observed in neutron scattering experiments.  In
these systems anisotropy vanishes in the critical limit, and $G(x)$
approaches a rotationally-invariant form.

We mention that the effects of a breakdown of ${\rm O}(N)$ symmetry 
in the interactions have been widely
considered in the literature~\cite{Zinn}.

Several approaches have been considered in order to study
the critical behavior of the two-point function 
\begin{equation}
G(x)=\langle \vec{s}(x) \cdot \vec{s}(0)\rangle.
\label{e5}
\end{equation}
In lattice ${\rm O}(N)$ non-linear $\sigma$ models, 
we have calculated the strong-coupling expansion of $G(x)$ 
up to 15th order
on the cubic lattice~\cite{nota2}
 and 21st order on the diamond lattice
within the corresponding nearest-neighbor formulations,
\begin{equation}
S_L = -N\beta \sum_{\rm links}  \vec{s}_{x_l}\cdot \vec{s}_{x_r},
\label{e6}
\end{equation}
where $x_l,x_r$ indicate the sites at the ends of each link.
Our conclusions are integrated and supported by results obtained
from a $1/N$ expansion of lattice ${\rm O}(N)$ models.
Furthermore, we have analyzed the first nontrivial term of 
the $\epsilon$-expansion and of the $g$-expansion (i.e.\ 
expansion in the coupling at fixed dimensions $d=3$) of
the two-point function within the corresponding $\phi^4$ formulation
of ${\rm O}(N)$ models.

Our results substantially confirm experimental observations that
scaling corrections to the Gaussian behavior (\ref{e4}) are very
small. This point has been already discussed in the literature,
essentially by $\epsilon$-expansion~\cite{Fisher} and $1/N$
expansion~\cite{Aharony} in the continuum formulation of ${\rm O}(N)$
models. We present here rather accurate determinations of such
corrections obtained by a strong-coupling analysis of $G(x)$.

We will show that the anisotropy of $G(x)$, for the class of systems
we are considering, vanishes at the rotationally-invariant fixed point,
with a behavior governed by a universal critical exponent $\rho$:
non-spherical moments (i.e.\ those which vanish when calculated on
spherical functions) of $G(x)$ are depressed with respect to spherical
moments carrying the same naive physical dimensions by a factor
$\xi^{-\rho}$.  From a field-theoretical point of view, anisotropy in
space is due to non rotationally-invariant (but ${\rm O}(N)$ symmetric)
irrelevant operators in the effective Hamiltonian, whose presence
depends essentially on the symmetries of the physical system, or of
the lattice formulation.  Another possible origin is the coupling of
the system to an external (anisotropic) source.  The exponent $\rho$
is related to the critical effective dimension of the leading
irrelevant operator breaking rotational invariance.  On cubic-like
lattices the leading operator has canonical dimension $d+2$.  In the
large-$N$ limit, where the canonical dimensions determine the scaling
properties, one then finds $\rho=2$.  $1/N$ expansion and
strong-coupling calculations show that, for all values of $N\geq 0$,
$\rho$ remains close to its canonical value.  One can also show that
for the two-dimensional Ising model the exact result $\rho=2$ holds.
Moreover, in the $1/N$ expansion of two-dimensional models, one finds
that, to $O(1/N)$, the multiplicative logarithmic correction to the
behavior $\sim\xi^{-2}$ is absent.

The technical details of our study will be reported in a separate
extended paper. Here we only describe the general features of our
analysis and the main numerical results.

For definiteness let us consider the cubic lattice version of the
models. We parametrize the two-point spin-spin
function by a multipole expansion  in the form 
\begin{equation}
\beta^{-1}\widetilde{G}^{-1}(k,M_G) = 
   \sum_{l=0}^\infty g_{2l}(y,M_G) Q_{2l}(k), 
\label{e7}
\end{equation}
where $y\equiv k^2/M_G^2$, and $Q_{2l}(k)$ are homogeneous functions
of momenta of degree $2l$ which are invariant under the symmetries of
the lattice. Their expressions can be obtained from the fully
symmetric traceless tensors of rank $2l$,
$T_{2l}^{\alpha_1\ldots\alpha_{2l}}(k)$, by considering all the
cubic-invariant combinations~\cite{nota1}.  Odd rank terms are absent
in the expansion (\ref{e7}) because of the parity symmetry. Moreover,
there is no rank-two term, due to the discrete rotational symmetry of
the lattice.  The first nontrivial function is
\begin{equation}
Q_{4}(k) =   \sum_\mu k_\mu^4 - {3\over 5} 
\Bigl(\sum_\mu k_\mu^2\Bigr)^2.
\label{e8}
\end{equation}
$Q_4(k)$  corresponds to the leading irrelevant operator
that breaks rotational invariance in the low-momentum
expansion of the Hamiltonian. In the continuum 
notation this operator has the form
\begin{equation}
O_4(x)=\vec{s}(x)\cdot Q_{4}(\partial)\vec{s}(x),
\label{e9}
\end{equation}
which has canonical dimension five in $d=3$.

The scaling limit of Eq.~(\ref{e7})
corresponds to  taking $M_G\rightarrow 0$ while keeping
the ratio $k/M_G$ finite. Hence we can parametrize the low-momentum
behavior of $G^{-1}(k,M_G)$ in the critical region by
\begin{eqnarray}
&&\beta^{-1}\widetilde{G}^{-1}(k,M_G)=
Z^{-1}\widehat{g}_0(y)M_G^2 +... \nonumber \\
&&\qquad+\, Z_4Q_4(k/M_G)\widehat{g}_4(y)M_G^4 +O(M_G^6),
\label{e10}
\end{eqnarray}
where we have dropped rotationally-invariant $O(M_G^4)$ terms 
which we are not interested in, and we have introduced the quantities
\begin{equation}
Z^{-1}={g_0(0,M_G) \over M_G^2},\qquad
Z_4= g_4(0,M_G),
\label{e11}
\end{equation}
which, in the limit $M_G\rightarrow0$, absorb all non-analytical
dependence on $M_G$ of the corresponding terms.  The functions
$\widehat{g}_0(y)$ and $\widehat{g}_4(y)$ are universal, i.e., they do
not depend on the specific form of the lattice hamiltonian. They
possess a regular expansion around $y=0$:
\begin{eqnarray}
\widehat{g}_0(y)&=& 1 + y + \sum_{i=2}^\infty c_i y^i, \\
\widehat{g}_4(y)&=& 1 + \sum_{i=1}^\infty d_i y^i .
\label{e12}
\end{eqnarray}

In the limit $N\rightarrow\infty$, the models are strictly Gaussian
and therefore all the coefficients $c_i$ and $d_i$ are zero,

We consider the spherical moments 
\begin{equation}
m_{2j}=\sum_x |x|^{2j}G(x),
\label{e11p3}
\end{equation}
and the leading non-spherical moments
\begin{equation}
q_{4,j} = \sum_x |x|^{2j} Q_4(x)G(x)
\label{e11p5}
\end{equation}
which vanish if  $G(x)$ is rotationally invariant.
The critical exponent $\rho$, describing the
vanishing of anisotropy, and the coefficients $c_i$ and $d_i$ of
the low-momentum expansion of $\widehat{g}_0(y)$ and $\widehat{g}_4(y)$, 
can be determined by studying appropriate combinations of the above moments
in the critical limit. 
In the critical region
\begin{equation}
{q_{4,m}\over m_{4+2m}} \sim {1\over \xi^{\rho}},
\label{crbeh}
\end{equation}
where $q_{2,m}$ and $m_{4+2m}$ have the same naive physical
dimensions.

In the case of non-Gaussian fixed points, like those corresponding to
the theory at finite $N$, $O_4(x)$ develops an anomalous dimension
$\sigma$, which causes a departure from the Gaussian value of $\rho$,
$\rho=2+\sigma$.  $\sigma$ can be extracted by evaluating the ratio
\begin{equation}
{ZZ_4\over M_G^2}={g_4(0,M_G)\over g_0(0,M_G)}\sim M_G^\sigma.
\label{sigmar}
\end{equation}
In turn this combination is easily estimated by taking
the moment ratio  $q_{4,0}/m_2\sim M_G^{\sigma}$.
We have calculated the first non-trivial term
of the $1/N$, $\epsilon$ and $g$-expansions
of $\sigma$:
\begin{equation}
\sigma={32\over 21\pi^2N}
+O\left({1\over N^2}\right),
\label{en19}
\end{equation}
\begin{equation}
\sigma={7\over20}\,{(N+2)\over(N+8)^2}\,\epsilon^2 
+O\left(\epsilon^3\right),
\label{en19b}
\end{equation}
\begin{equation}
\sigma= \widetilde{g}^{*2}\,{5408\over25515}\,{(N+2)\over (N+8)^2}
+ O\left(\widetilde{g}^{*3}\right)
\label{en19c}
\end{equation}
($\widetilde{g}^*$ is the fixed-point value of the rescaled coupling:
$\widetilde{g}=g(N+8)/48\pi$~\cite{Zinn}).  Strong-coupling estimates
of $\sigma$ have been obtained by analyzing and comparing the
available strong-coupling series of $q_{4,0}$ and $m_2$ on both cubic
and diamond lattices.  Universality between cubic and diamond lattice
is substantially verified, although the analysis on the diamond
lattice leads to less precise results. In Table~\ref{sigma} we report
the strong-coupling estimates of $\sigma$, and for comparison its
estimates from Eqs.~(\ref{en19}), (\ref{en19b}) and (\ref{en19c}).
For ${\rm O}(N)$ models, the values of $\sigma$ are very small in the
whole range $N\geq 0$, thus indicating an essentially Gaussian
behavior of this critical exponent. The comparison of the estimates of
$\sigma$ from the various approaches we have considered is
satisfactory.

The low-momentum critical behavior of the two-point 
Green's function, i.e. 
\begin{equation}
{\widetilde{G}(0)\over \widetilde{G}(k)} \rightarrow \widehat{g}_0(y),
\end{equation}
has been already investigated within an $\epsilon$-expansion of the
$\phi^4$ formulations of ${\rm O}(N)$ models (up to
$O(\epsilon^2)$)~\cite{Fisher}, and in the $1/N$ expansion (up to
$O(1/N)$)~\cite{Aharony}.  The computations of the $O(1/N^2)$
contribution to $\widehat{g}_0(y)$ is in progress.  Moreover rough
estimates of such corrections have been presented for $N=1$ and $N=3$
by analyzing strong-coupling expansion (up to 10th order) within
various lattice formulations of the theory~\cite{Fisher}.  We have
reconsidered the problem of determining the non-Gaussian corrections
to $\widehat{g}_0(y)$ in the low-momentum regime, and those of
$\widehat{g}_4(y)$, especially by a strong-coupling analysis.
Furthermore we have calculated the first nontrivial contributions to
$\widehat{g}_4(y)$ in the $1/N$, $\epsilon$ and $g$-expansions, and of
$\widehat{g}_0(y)$ in the $g$-expansion, and the corresponding
coefficients of the expansion around $y=0$.

Our strong-coupling analysis leads to a satisfactory
precision which, in the case of the coefficient $c_i$, 
considerably improves earlier calculations.
This has been achieved essentially for two reasons:
longer strong-coupling series are available, and,
even more important,    improved estimators have been employed.
We indeed took special care in the choice of estimators for the physical
quantities $c_i$ and $d_i$. This is very important from a
practical point view: better estimators can greatly improve the
stability of the extrapolation to the critical point. Our search for
optimal estimators was guided by the knowledge of the large-$N$ limit.
We chose estimators which are perfect for $N=\infty$, i.e.\ do not
present off-critical corrections to their critical value
$c_i=d_i=0$ in the symmetric phase.
In particular in the case of $c_2$ and $d_1$ and  on the cubic
lattice, we used the following estimators:
\begin{eqnarray}
e(c_2)&=& 1 - {m_4\over 120 m_0}\,M_G^4+ {1\over 20}\,M_G^2,\\
e(d_1)&=& 2 \left( 1 - {q_{4,1}\over 44 q_{4,0}}\,M_G^2 + 
        {1\over 44}\,M_G^2\right)
\end{eqnarray}
respectively.  In Table~\ref{summary} we report our strong-coupling
estimates for $c_2$, $c_3$ and $d_1$ on both cubic and diamond
lattice, obtained by evaluating at $\beta_c$ appropriate approximants
(such as Pad\'e and first order integral approximants) of the
strong-coupling series of the corresponding estimators.  One may
notice that universality between cubic and diamond lattice is always
confirmed.  At large-$N$ the strong-coupling estimates of $c_2$, $c_3$
and $d_1$ compare very well with the corresponding $O(1/N)$ formulas:
$c_2=-0.004449/N+...$, $c_3=0.000134/N+...$, etc...\cite{Aharony},
$d_1=-0.0020647/N+...$, $d_2=0.0000738/N+...$, etc....  The comparison
with the estimates from other approaches, which are reported in
Table~\ref{summary}, is good.

All calculations agree in indicating that the inequality
\begin{equation}
c_i\ll c_2\ll 1 \qquad {\rm for}\quad i\geq 3
\label{pattern}
\end{equation}
is satisfied for all $N$.
These results show that, in the critical
region of the symmetric phase, the two-point Green's function
is substantially Gaussian
in a large region around $k^2$, i.e.\ for $|k^2/M_G^2|\lesssim 1$,
for all $N$ from zero to infinity.  
Another important consequence of the relation (\ref{pattern}) 
is the possibility of
evaluating the zero of $\widehat{g}_0(y)$ closest to the origin,
$y_0$, to a rather good approximation by the relationship 
\begin{equation}
-y_0\simeq 1+c_2. 
\label{y0}
\end{equation}
The quantity $-y_0$ in turn is the scaling limit of the ratio of the
correlation length defined in Eq.~(\ref{e3}) with the ``true''
correlation length obtained from the damping factor in the exponential
long-distance behavior of $G(x)$.  Direct calculations of this ratio
confirm the relationship (\ref{y0}).  Similar results have been
obtained in the two-dimensional ${\rm O}(N)$
models~\cite{ON-d2-a,ON-d2-b}.

In general models defined on non-Bravais lattices such as the diamond
lattice are not parity-invariant, and odd-rank operators are allowed
in the corresponding expansion of the effective Hamiltonian.  We
finally discuss how space-parity violating terms, when they exist,
vanish approaching the rotationally-invariant fixed point.  This fact
should be described by a critical exponent $\rho_p$ which should be
universal in systems breaking parity at a microscopic level, such as
ferromagnetic materials having the structure of a diamond lattice.
$\rho_p$ can be evaluated on the nearest-neighbor formulation of ${\rm
O}(N)$ models on the diamond lattice, which is not parity invariant.
In the correspoding Gaussian theory, or the large-$N$ limit of ${\rm
O}(N)$ models, one has $\rho_p=3$, indeed one finds that
\begin{equation}
{\sum xyz \, G(x,y,z)\over \sum G(x,y,z)}
\longrightarrow {\rm const.}
\end{equation}
in the critical limit.  In general, for finite $N$, $\rho_p$ may
differ from its Gaussian value.  The strong-coupling analysis of the
odd moments of $G(x)$ on the diamond lattice shows that the correction
to the Gaussian value of $\rho_p$ is very small. We estimated that
$0\leq \rho_p-3\lesssim 0.01$ for all $N\geq 0$.


\newpage
\begin{table}
\squeezetable
\caption{
For various values of $N$, we report the estimates of $\sigma$
obtained by our strong-coupling analysis, and by the first nontrivial
terms in the $1/N$ expansion, $\epsilon$-expansion, and $g$-expansion.
The errors diplayed in the strong-coupling estimates are a rough
estimate of the uncertainty, which takes into account all the analyses
we have performed.}
\label{sigma}
\begin{tabular}{cr@{}lr@{}lr@{}lr@{}l}
\multicolumn{1}{c}{$N$}&
\multicolumn{2}{c}{s.c.}&
\multicolumn{2}{c}{$1/N$ exp}&
\multicolumn{2}{c}{$\epsilon$-exp}&
\multicolumn{2}{c}{$g$-exp}\\
\tableline \hline
0 & 0&.00(1) & &    & 0&.0109 &  0&.0134 \\
1 & 0&.01(1) & &    & 0&.0130 &  0&.0157 \\
2 & 0&.02(1) & &    & 0&.0140 &  0&.0168 \\
3 & 0&.03(2) & 0&.0515 & 0&.0145 & 0&.0170 \\
4 & 0&.03(2) & 0&.0386 & 0&.0147 & 0&.0165 \\
8 & 0&.02(1) & 0&.0193 & 0&.0137 & 0&.0116 \\
16 & 0&.009(3)& 0&.0096 & 0&.0109 & 0&.0096\\
32 & 0&.004(2)& 0&.0048 & 0&.0074 & 0&.0054 \\
\end{tabular}
\end{table}

\begin{table}
\squeezetable
\caption{
Strong-coupling estimates of the coefficients $c_2$ and $c_3$ of the
expansion of $\widehat{g}_0(y)$, and the coefficient $d_1$ of the
expansion of $\widehat{g}_4(y)$.  For comparison, we also report
estimates from the first nontrivial terms in the $\epsilon$-expansion,
$g$-expansion, and $1/N$ expansion.  The errors diplayed in the
strong-coupling estimates should give an idea of the spread of the
results from the various Pad\'e$-$type and integral approximants we
considered in our analysis.}
\label{summary}
\begin{tabular}{ccr@{}lr@{}lr@{}lr@{}l}
\multicolumn{1}{c}{$N$}&
\multicolumn{1}{c}{}&
\multicolumn{2}{c}{$c_2$}&
\multicolumn{2}{c}{$c_3$}&
\multicolumn{2}{c}{$d_1$}\\
\tableline \hline
0 & cubic   & $|c_2|\lesssim$ 2& $\times10^{-4}$ & 1&.2(1)$\times10^{-5}$ & 
1&(1)$\times10^{-4}$ \\
  & diamond & $|c_2|\lesssim$ 1&$\times 10^{-4}$ & 1&.0(1)$\times10^{-5}$ & 
$-$1&.0(5)$\times10^{-4}$ \\
  & $\epsilon$-exp. & $-$2&.35 $\times10^{-4}$ & 0&.60$\times10^{-5}$ &
$-$1&.11$\times10^{-4}$ \\
  & $g$-exp. & $-$3&.12$\times10^{-4}$ & 1&.65$\times10^{-5}$ & 
$-$1&.42$\times10^{-4}$ \\\hline

1 & cubic   & $-$2&.9(2)$\times10^{-4}$ & 1&.1(1)$\times10^{-5}$ & 
$-$1&.7(5)$\times10^{-4}$ \\
  & diamond & $-$3&.1(2)$\times10^{-4}$ & 1&.0(2)$\times10^{-5}$ & 
$-$3&(1)$\times10^{-4}$ \\
  & $\epsilon$-exp. & $-$2&.78 $\times10^{-4}$ & 0&.71$\times10^{-5}$ & 
$-$1&.31$\times10^{-4}$ \\
  & $g$-exp. & $-$3&.66$\times10^{-4}$ & 1&.94$\times10^{-5}$ & 
$-$1&.67$\times10^{-4}$   \\\hline

2 & cubic   & $-$3&.8(3)$\times10^{-4}$ & 1&.1(1)$\times10^{-5}$ & 
$-$2&.3(2)$\times10^{-4}$ \\
  & diamond & $-$4&.2(3)$\times10^{-4}$ & 1&.1(3)$\times10^{-5}$ & 
$-$3&(1)$\times10^{-4}$ \\
  & $\epsilon$-exp. & $-$3&.01$\times10^{-4}$ & 0&.77$\times10^{-5}$ & 
$-$1&.42$\times10^{-4}$ \\
  & $g$-exp. & $-$3&.90$\times10^{-4}$ & 2&.06$\times10^{-5}$ & 
$-$1&.78$\times10^{-4}$   \\\hline

3 & cubic   & $-$4&.0(2)$\times10^{-4}$& 1&.1(2)$\times10^{-5}$ & 
$-$2&.5(2)$\times10^{-4}$ \\
  & diamond & $-$4&.6(4)$\times10^{-4}$& 1&.1(3)$\times10^{-5}$ & 
$-$2&.6(3)$\times10^{-4}$ \\
  & $\epsilon$-exp. & $-$3&.11$\times10^{-4}$ & 0&.79$\times10^{-5}$ & 
$-$1&.46$\times10^{-4}$ \\
  & $g$-exp. & $-$3&.95$\times10^{-4}$ & 2&.09$\times10^{-5}$ & 
$-$1&.80$\times10^{-4}$  \\\hline

4 & cubic   & $-$4&.1(2)$\times10^{-4}$ & 1&.2(1)$\times10^{-5}$ & 
$-$2&.5(2)$\times10^{-4}$ \\
  & diamond & $-$4&.7(2)$\times10^{-4}$ & 1&.0(2)$\times10^{-5}$ & 
$-$2&.5(5)$\times10^{-4}$ \\
  & $\epsilon$-exp. & $-$3&.13$\times10^{-4}$ & 0&.80$\times10^{-5}$ & 
$-$1&.48$\times10^{-4}$ \\
  & $g$-exp. & $-$3&.85 $\times10^{-4}$ & 2&.04$\times10^{-5}$ & 
$-$1&.76$\times10^{-4}$  \\
  & $1/N$ exp.     & $-$11&.12$\times10^{-4}$ & 3&.36$\times10^{-5}$ & 
$-$5&.16$\times10^{-4}$ \\\hline

8 & cubic   & $-$3&.5(2)$\times10^{-4}$& 1&.0(2)$\times10^{-5}$ & 
$-$2&.1(2)$\times10^{-4}$ \\
   & diamond & $-$4&.0(1)$\times10^{-4}$& 0&.7(5)$\times10^{-5}$ & 
$-$3&(2)$\times10^{-4}$ \\
  & $\epsilon$-exp. & $-$2&.94$\times10^{-4}$& 0&.75$\times10^{-5}$ & 
$-$1&.38$\times10^{-4}$ \\
  & $g$-exp. & $-$2&.70 $\times10^{-4}$ & 1&.43$\times10^{-5}$ & 
$-$1&.24$\times10^{-4}$  \\
  & $1/N$ exp.     & $-$5&.56  $\times10^{-4}$&1&.18$\times10^{-5}$ & 
$-$2&.58$\times10^{-4}$  \\\hline

16& cubic   & $-$2&.4(2)$\times10^{-4}$& 0&.70(5)$\times10^{-5}$& 
$-$1&.4(2)$\times10^{-4}$ \\
  & diamond & $-$2&.65(5)$\times10^{-4}$& 0&.5(3)$\times10^{-5}$ & 
$-$1&.2(8)$\times10^{-4}$ \\
  & $\epsilon$-exp. & $-$2&.35$\times10^{-4}$& 0&.60$\times10^{-5}$ &
$-$1&.11$\times10^{-4}$ \\
  & $g$-exp. & $-$2&.25 $\times10^{-4}$ & 1&.18$\times10^{-5}$ & 
$-$1&.03$\times10^{-4}$  \\
  & $1/N$ exp.     & $-$2&.78$\times10^{-4}$& 0&.84$\times10^{-5}$ & 
$-$1&.29$\times10^{-4}$ \\\hline

32& cubic   & $-$1&.44(5)$\times10^{-4}$& 0&.42(4)$\times10^{-5}$& 
$-$0&.7(2)$\times10^{-4}$ \\
  & diamond & $-$1&.50(5)$\times10^{-4}$ & 0&.3(2)$\times10^{-5}$ & 
$-$0&.5(3)$\times10^{-4}$ \\
  & $\epsilon$-exp. & $-$1&.60$\times10^{-4}$ & 0&.41$\times10^{-5}$ & 
$-$0&.75$\times10^{-4}$ \\
  & $g$-exp. & $-$1&.25$\times10^{-4}$ & 0&.66$\times10^{-5}$ & 
$-$0&.57$\times10^{-4}$  \\
  & $1/N$ exp.     & $-$1&.39$\times10^{-4}$& 0&.42$\times10^{-5}$ & 
$-$0&.64$\times10^{-4}$ \\\hline
$\infty$ &  & 0 &                           
  & 0  &                & 0&
\end{tabular}
\end{table}


\begin{references}
\bibitem{nota1}
For given rank $2l$, there may be more than one invariant combination when 
$2l\geq 12$ in three dimensions. Such a degeneracy is irrelevant to our
discussion.
\end{references}
\end{document}